\documentclass[aps,pra,superscriptaddress,reprint,floatfix]{revtex4-2}

\usepackage{amsmath,amssymb,amsthm,mathtools}
\usepackage[hidelinks]{hyperref}
\usepackage{braket}
\usepackage{graphicx}

\newtheorem{theorem}{Theorem}

\newcommand{\bbE}{\mathbb{E}}
\newcommand{\bbP}{\mathbb{P}}
\newcommand{\kB}{k_{\mathrm B}}
\newcommand{\Id}{\mathbb I}

\begin{document}

\title{Thermodynamic Value of XOR-Game-Induced Side Information in a Szilard Engine: Extremal Binary-Channel Bounds}

\author{Piotr {\'C}wikli{\'n}ski}
\affiliation{International Centre for Theory of Quantum Technologies,
University of Gda{\'n}sk, ul. prof. Marii Janion 7, 80-309 Gda\'nsk, Poland}

\date{\today}

\begin{abstract}
In this work, we study the reversible work value of a binary
side-information channel supplied to a Szilard engine. The working system is
a classical thermal bit, while the feedback controller receives only a binary
record correlated with its physical state. We show that a fixed average
prediction accuracy does not determine a unique thermodynamic value, because
different binary channels with the same accuracy can carry different amounts
of mutual information. We derive the exact minimum and maximum reversible
Szilard work compatible with a given prediction accuracy. The minimum is
attained uniquely by the binary symmetric channel, whereas the maximum is
attained by two asymmetric channels for which all errors are concentrated on
one input value. We then show that the standard XOR-game embedding always
produces the symmetric channel and therefore realizes the minimum possible
Szilard work among all binary side-information channels with the same
prediction accuracy. Finally, we contrast this information-to-work valuation
with an event-routing battery valuation, which depends only on the task-success
probability.
\end{abstract}

\maketitle

\section{Introduction}
\label{sec:introduction}

The connection between information and thermodynamics goes back to Maxwell,
Szilard, and the later analysis of information erasure by Landauer and Bennett
\cite{Szilard1929,Landauer1961,Bennett1982,MaruyamaNoriVedral2009}.
Modern feedback thermodynamics makes this connection quantitative:
information about a physical system can increase the amount of work extracted
from a heat bath, while closing the information cycle carries its own
thermodynamic cost
\cite{SagawaUeda2008,SagawaUeda2010,Parrondo2015,Goold2016,
deOliveira2025Tutorial}.  Information-to-energy conversion and Landauer
erasure have also been demonstrated experimentally
\cite{Toyabe2010InformationEnergy,Koski2014SingleElectronSzilard,
Berut2012LandauerExperiment}.

Let us first describe the physical situation considered here.  The working
medium is a classical two-state system \(S\), for example a single particle in
a Szilard box with left and right cells.  The system is initially in thermal
equilibrium with a heat bath at temperature \(T\), the two logical states have
equal energy, and its physical state is represented by a uniformly distributed
bit \(X\).  An external information-processing stage produces a binary record
\(G\).  The feedback controller of the Szilard engine receives only \(G\) and
performs a reversible feedback protocol on \(S\).

The quantity directly relevant for the reversible work is the mutual
information \(I(X{:}G)\).  A natural operational summary of the record is its
prediction accuracy
\[
p:=\bbP[G=X].
\]
However, in contrast with a binary symmetric channel, a general binary channel
is not determined by this single number.  Two records can have the same
prediction accuracy and nevertheless carry different amounts of information
about \(X\).  This immediately raises the question considered in this paper:
for a fixed \(p\), what is the smallest and the largest reversible work value
that a binary side-information channel can have?

We answer this question exactly.  The minimum is
\[
\kB T\ln2\,[1-h_2(p)],
\]
and is attained uniquely, for \(1/2<p<1\), by the binary symmetric channel.
The maximum is
\[
\kB T\ln2
\left[
h_2\!\left(p-\frac12\right)
-\frac12h_2(2p-1)
\right],
\]
and is attained by the two channels for which one input symbol is transmitted
without error and all errors are concentrated on the other input symbol.
Thus, a binary prediction score determines an interval of thermodynamic
values, rather than a unique one.

This observation is useful when the external information-processing stage is
built from a Bell or XOR game.  Bell nonlocality distinguishes local, quantum,
and more general nonsignalling correlations
\cite{Bell1964,CHSH1969,Tsirelson1980,PopescuRohrlich1994,
BarrettPRA2005,BrunnerRMP2014}.  XOR games provide a convenient language for
binary-output Bell tasks
\cite{CleveHoyerTonerWatrous2004,BraunsteinCaves1990,Wehner2006}.  In the
standard XOR embedding, the error bit of the induced record is independent of
the thermal bit \(X\).  The resulting channel is therefore binary symmetric.
Consequently, the XOR construction does not merely lead to the familiar
formula \(1-h_2(\omega_{\mathcal G})\): it realizes the lower thermodynamic
envelope among all binary channels with the same prediction accuracy
\(p=\omega_{\mathcal G}\).

The restricted character of the construction should be kept in mind.  The
Szilard engine values the effective channel \(P(G|X)\), not the microscopic
origin of that channel.  Two physically different upstream mechanisms which
produce the same \(P(G|X)\) have the same value in the present model.  Thus the
construction does not provide a new Bell bound or a device-independent
thermodynamic witness.  Work extraction from correlations, discord, steering,
and other quantum resources has been studied in different physical settings
\cite{OppenheimHorodecki2002WorkDeficit,Zurek2003DiscordDemon,
Francica2017DaemonicErgotropy,ManzanoPlastinaZambrini2018,
PerarnauLlobet2015,Bera2017,BeyerLuomaStrunz2019,
BiswasSteeringWork2025,Ji2022,deOliveira2025Heat,HsiehGessner2026}.
The present problem is instead a channel-level information-to-work valuation.

A second motivation comes from the battery-explicit construction of
Ref.~\cite{cwiklinski2026battery_witness}.  There, a successful game round
routes a pre-supplied excitation to a battery, so the mean energetic response
depends only on the success probability.  In the Szilard setting, by contrast,
the work is proportional to \(I(X{:}G)\), and the same success probability can
correspond to a whole interval of work values.  The XOR embedding occupies one
extremal boundary of this interval.  This comparison makes explicit that an
abstract task score does not possess a unique thermodynamic value independently
of the physical mechanism in which it is used.

This work is organized as follows.  In Sec.~\ref{sec:szilard}, we introduce
the Szilard working system and recall the reversible value of a supplied
side-information channel.  In Sec.~\ref{sec:extrema}, we derive the exact
thermodynamic interval at fixed binary prediction accuracy.  In
Sec.~\ref{sec:xor}, we show that the XOR-game construction realizes the lower
boundary and discuss CHSH as an example.  Section~\ref{sec:battery} compares
the channel valuation with the battery-routing valuation.  The scope of the
thermodynamic bookkeeping is discussed in Sec.~\ref{sec:bookkeeping}, and we
conclude in Sec.~\ref{sec:discussion}.

\section{Szilard value of a binary side-information channel}
\label{sec:szilard}

We consider a classical two-state working system with pointer basis
\(\{\ket{0},\ket{1}\}\).  The two states can be realized by the left and right
cells of a one-particle Szilard box.  The initial Hamiltonian is degenerate,
\begin{equation}
H_S^{(0)}=0,
\label{eq:H0}
\end{equation}
and the system is equilibrated with a bath at inverse temperature
\(\beta=(\kB T)^{-1}\).  Hence the physical bit \(X\) is uniform:
\[
\bbP[X=0]=\bbP[X=1]=\frac12.
\]
The preferred classical basis is part of the physical model; it is not selected
by the degenerate Hamiltonian itself.

Suppose that the feedback controller receives a binary record \(G\).
Conditioned on \(G=g\), the state of the working system is
\[
\rho_{S|g}
=
\sum_x P(x|g)\ket{x}\!\bra{x}.
\]
For the degenerate two-level system, the nonequilibrium free-energy excess of
this conditional state over the equilibrium state \(\Id/2\) is
\[
\Delta F_g
=
\kB T\ln2\,
D_2\!\left(P(X|g)\middle\|U_2\right),
\]
where \(U_2=(1/2,1/2)\).  In the ideal reversible feedback limit, this
free-energy difference can be extracted as work
\cite{SagawaUeda2008,SagawaUeda2010,Parrondo2015}.  Averaging over \(g\)
gives
\begin{align}
W_{\rm Sz}(X{:}G)
&=
\kB T\ln2
\sum_g P(g)
D_2\!\left(P(X|g)\middle\|P(X)\right)
\nonumber\\
&=
\kB T\ln2\,I(X{:}G).
\label{eq:info_work}
\end{align}

The same expression follows directly in the one-particle Szilard picture.
Conditioned on \(G=g\), the partition is moved quasistatically so that
\[
V_x^{(g)}=P(X=x|G=g)V.
\]
If the particle is in cell \(x\), the reversible isothermal work is
\[
W(x|g)
=
\kB T\ln\!\left[2P(X=x|G=g)\right].
\]
Averaging over \(x\) and \(g\) reproduces Eq.~\eqref{eq:info_work}.  Thus the
thermodynamic question at fixed prediction accuracy reduces to a purely
information-theoretic extremal problem for binary channels.

\section{Exact work interval at fixed prediction accuracy}
\label{sec:extrema}

Let
\begin{equation}
e_0:=\bbP[G=1|X=0],
\qquad
e_1:=\bbP[G=0|X=1]
\label{eq:error_probs}
\end{equation}
be the two conditional error probabilities.  Since \(X\) is uniform,
\begin{equation}
p
=
\bbP[G=X]
=
1-\frac{e_0+e_1}{2}.
\label{eq:accuracy}
\end{equation}
We orient the output labels so that \(p\ge1/2\).  Put
\[
q:=1-p
\]
and parametrize all binary channels with the same prediction accuracy as
\begin{equation}
e_0=q+t,
\qquad
e_1=q-t,
\qquad
-q\le t\le q.
\label{eq:tparam}
\end{equation}
The parameter \(t\) measures the asymmetry between the two conditional error
probabilities.

A direct calculation gives
\begin{equation}
I_p(t)
=
h_2\!\left(\frac12+t\right)
-\frac12 h_2(q+t)
-\frac12 h_2(q-t).
\label{eq:Ipt}
\end{equation}
The main result of this paper is the following.

\begin{theorem}[Binary-channel Szilard interval]
\label{thm:extrema}
Let \(X\) be uniform on \(\{0,1\}\), let \(G\in\{0,1\}\), and fix
\[
p=\bbP[G=X]\in[1/2,1].
\]
Then the ideal reversible Szilard work satisfies
\begin{equation}
W_{\min}(p)
\le
W_{\rm Sz}(X{:}G)
\le
W_{\max}(p),
\label{eq:work_interval}
\end{equation}
where
\begin{align}
W_{\min}(p)
&=
\kB T\ln2\,[1-h_2(p)],
\label{eq:wmin}\\
W_{\max}(p)
&=
\kB T\ln2
\left[
h_2\!\left(p-\frac12\right)
-\frac12h_2(2p-1)
\right].
\label{eq:wmax}
\end{align}
For \(1/2<p<1\), the lower bound is attained uniquely by the binary symmetric
channel
\begin{equation}
e_0=e_1=1-p,
\label{eq:bscmin}
\end{equation}
whereas the upper bound is attained exactly by the two \(Z\)-type channels
\begin{equation}
(e_0,e_1)
=
\left(0,2(1-p)\right)
\quad\text{or}\quad
\left(2(1-p),0\right).
\label{eq:zmax}
\end{equation}
\end{theorem}

The proof is given in Appendix~\ref{app:proof}.  The important point is that
the function \(I_p(t)\) is even and strictly convex for \(1/2<p<1\).
Consequently, its unique minimum is at \(t=0\), whereas its maxima are the two
endpoints \(t=\pm q\).

\begin{figure}[t]
\centering
\includegraphics[width=\columnwidth]{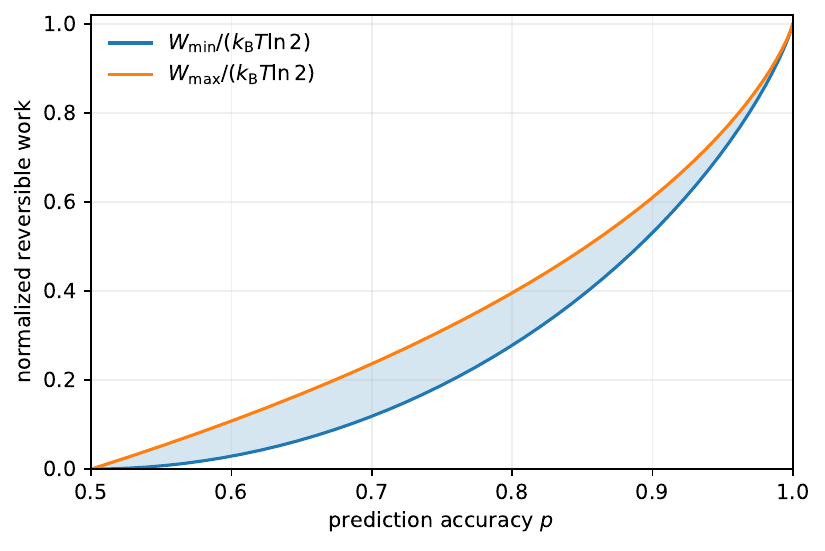}
\caption{
Normalized reversible Szilard work at fixed prediction accuracy
\(p=\Pr[G=X]\).  The lower boundary is the binary symmetric channel,
\(W_{\min}/(\kB T\ln2)=1-h_2(p)\).  The upper boundary is attained by the two
\(Z\)-type channels.  Every binary side-information channel with uniform input
lies in the shaded region.
}
\label{fig:work_region}
\end{figure}

Figure~\ref{fig:work_region} shows the normalized interval.  At fixed
prediction accuracy, the shaded region consists of all reversible Szilard
values compatible with a binary record.  The lower boundary corresponds to the
binary symmetric channel, and the upper boundary to the two \(Z\)-type
channels.

The difference between the two boundaries is already visible close to random
guessing.  If
\[
p=\frac12+\delta,
\qquad
\delta\to0^+,
\]
then
\begin{align}
\frac{W_{\min}}{\kB T\ln2}
&=
\frac{2\delta^2}{\ln2}
+O(\delta^4),
\label{eq:wminexp}\\
\frac{W_{\max}}{\kB T\ln2}
&=
\delta+\frac{\delta^2}{2\ln2}
+O(\delta^3).
\label{eq:wmaxexp}
\end{align}
Thus records with the same small excess prediction accuracy can have
parametrically different thermodynamic values: quadratic for the symmetric
channel and linear for the extremal asymmetric channels.

\section{XOR games as the lower thermodynamic envelope}
\label{sec:xor}

A finite two-player XOR game is specified by
\[
\mathcal G=(\mathcal U,\mathcal V,\mu,f),
\]
where \(\mathcal U\) and \(\mathcal V\) are the question sets,
\(\mu(u,v)\) is the input distribution, and
\(f(u,v)\in\{0,1\}\) is the winning predicate
\cite{CleveHoyerTonerWatrous2004}.  Alice and Bob receive \(u\) and \(v\) and
return bits \(a,b\).  They win when
\[
a\oplus b=f(u,v).
\]
A behaviour \(P(a,b|u,v)\) wins with probability
\begin{equation}
\omega_{\mathcal G}(P)
=
\sum_{u,v}\mu(u,v)
\sum_{a\oplus b=f(u,v)}
P(a,b|u,v).
\label{eq:omega}
\end{equation}

The physical Szilard bit \(X\) is sampled independently of the game.  The
external preprocessing stage forms
\begin{equation}
r=X\oplus f(u,v),
\qquad
G=a\oplus b\oplus r,
\label{eq:xorencoding}
\end{equation}
and only \(G\) is supplied to the feedback controller.  The corresponding
error bit is
\begin{equation}
G\oplus X
=
a\oplus b\oplus f(u,v).
\label{eq:xorerror}
\end{equation}
Because the sampled questions and the behaviour \(P(a,b|u,v)\) are independent
of \(X\), the distribution of this error bit is independent of \(X\).
Therefore the induced channel is binary symmetric, with
\begin{equation}
\bbP[G=X]
=
\omega_{\mathcal G}(P).
\label{eq:xoraccuracy}
\end{equation}

Theorem~\ref{thm:extrema} then gives a stronger interpretation of the XOR
formula than the channel calculation alone.  For every
\(\omega_{\mathcal G}(P)\in(1/2,1)\),
\begin{equation}
W_{\rm Sz}^{\mathcal G}(P)
=
W_{\min}\!\left(\omega_{\mathcal G}(P)\right)
=
\kB T\ln2
\left[
1-h_2\!\left(\omega_{\mathcal G}(P)\right)
\right].
\label{eq:xorwork}
\end{equation}
Hence the XOR embedding realizes the unique minimum-work binary channel among
all records with the same prediction accuracy.  The game score fixes the work
only because the XOR construction fixes the channel symmetry.

The accessible-record restriction remains essential.  The feedback controller
receives \(G\), but not \(u,v,r\) separately.  If it had access to those
variables, it could reconstruct \(X=r\oplus f(u,v)\) without using the game
outputs.  Thus Eq.~\eqref{eq:xorwork} is a valuation of the supplied effective
channel, not the net work of a fully autonomous Bell experiment.

For CHSH, the inputs are uniform, \(u,v\in\{0,1\}\), and \(f(u,v)=uv\).
Writing
\[
S=E_{00}+E_{01}+E_{10}-E_{11},
\]
one has
\[
\omega_{\rm CHSH}
=
\frac12+\frac{S}{8}.
\]
Equation~\eqref{eq:xorwork} therefore becomes
\begin{equation}
W_{\rm Sz}^{\rm CHSH}(S)
=
\kB T\ln2
\left[
1-h_2\!\left(\frac12+\frac{S}{8}\right)
\right].
\label{eq:chshwork}
\end{equation}
Inserting the known local, quantum, and nonsignalling values
\[
S_{\mathsf L}=2,\qquad
S_{\mathsf Q}=2\sqrt2,\qquad
S_{\mathsf{NS}}=4
\]
\cite{CHSH1969,Tsirelson1980,PopescuRohrlich1994} gives
\begin{align}
\frac{W_{\mathsf L}}{\kB T\ln2}
&\simeq0.1887,
\nonumber\\
\frac{W_{\mathsf Q}}{\kB T\ln2}
&\simeq0.3991,
\nonumber\\
\frac{W_{\mathsf{NS}}}{\kB T\ln2}
&=1.
\label{eq:chshvalues}
\end{align}
These values are not new Bell bounds.  They are the images of the standard
CHSH values under the lower-envelope Szilard valuation.

\section{Event routing versus channel valuation}
\label{sec:battery}

The distinction between prediction accuracy and full channel information is
particularly transparent when compared with the battery-explicit construction
of Ref.~\cite{cwiklinski2026battery_witness}.  In that protocol, a supplied
excitation of energy \(\Delta\) is routed to a battery when the external task
succeeds.  For a task-success probability \(p\),
\begin{equation}
\bbE[W_{\rm bat}]
=
\Delta p.
\label{eq:battery}
\end{equation}
Thus the energetic response is fixed once \(p\) is fixed.

The Szilard value is different.  At the same prediction accuracy \(p\),
Theorem~\ref{thm:extrema} allows the full interval
\[
W_{\min}(p)
\le
W_{\rm Sz}
\le
W_{\max}(p).
\]
The difference is physical.  Event routing assigns a fixed energetic value to
a successful event.  Szilard feedback values the information contained in the
record, and this depends on how errors are distributed conditionally on the
input.  Concentrating all errors on one value of \(X\) leaves more information
than distributing them symmetrically, even though the average number of
correct predictions is the same.

The XOR embedding occupies the symmetric lower boundary.  Therefore the
comparison with the battery construction is not simply between an affine and a
nonlinear function of the same scalar score.  More generally, the battery
valuation depends only on \(p\), whereas the Szilard valuation depends on the
whole channel \(P(G|X)\).  The XOR algebra removes this extra freedom by forcing
the binary symmetric channel.

\section{Thermodynamic scope and bookkeeping}
\label{sec:bookkeeping}

The boundary of the present model is intentionally narrow.  The working system
\(S\), the heat bath, and the controller record \(G\) form the effective
feedback module.  The referee, question generation, the devices producing
\(P(a,b|u,v)\), communication of their outputs, and the preprocessing required
to construct \(G\) are external.  Consequently, Eq.~\eqref{eq:xorwork} should
not be interpreted as the net work of a fully autonomous implementation.

If the controller memory containing \(G\) is reset blindly, Landauer's
principle gives
\[
W_{\rm reset}\ge\kB T\ln2\,H(G).
\]
The feedback stage yields at most
\(\kB T\ln2\,I(X{:}G)\), and hence the minimal module obeys
\begin{align}
W_{\rm fb}-W_{\rm reset}
&\le
\kB T\ln2[I(X{:}G)-H(G)]
\nonumber\\
&=
-\kB T\ln2\,H(G|X)
\le0.
\label{eq:reset}
\end{align}
If further variables such as \(u,v,r,a,b\) are retained as physical memories,
their reset or reversible uncomputation has to be accounted for separately.
Equation~\eqref{eq:reset} is therefore the bookkeeping of the minimal feedback
module, not of the complete task-generating apparatus.

The ideal work in Eq.~\eqref{eq:info_work} also assumes reversible feedback.
For a finite-time implementation,
\[
W_{\rm ext}
=
W_{\rm rev}-\kB T\,\Sigma
\]
when the dimensionless entropy production \(\Sigma\ge0\) is defined by this
identity.  Slow-driving corrections can be treated using, for example,
thermodynamic-friction methods
\cite{SivakCrooks2012,ZulkowskiDeWeese2015}.

\section{Discussion}
\label{sec:discussion}

In this work, we have determined the exact reversible Szilard-work interval
compatible with a fixed binary prediction accuracy.  A single success
probability does not determine the thermodynamic value of a general binary
side-information channel.  For \(1/2<p<1\), the binary symmetric channel is
the unique minimum-work channel, whereas two asymmetric \(Z\)-type channels
maximize the work.

The XOR-game construction acquires a simple interpretation in this general
picture.  Its error bit is independent of the thermal microstate, and it
therefore produces the binary symmetric channel.  Consequently, for a given
XOR-game score, the familiar value
\[
\kB T\ln2[1-h_2(\omega_{\mathcal G})]
\]
is the minimum Szilard value compatible with that prediction accuracy, not an
arbitrary consequence of choosing a symmetric example.

The result remains a channel-level thermodynamic statement.  The Szilard
module is insensitive to whether a given \(P(G|X)\) was produced by local,
quantum, post-quantum, or classical upstream hardware.  Known Bell bounds
enter only through the corresponding task scores.  What is new is the
information-thermodynamic structure at fixed score and the extremal role of
the XOR embedding within that structure.

Finally, the comparison with the battery-routing construction shows that the
physical embedding of a task score matters.  Event routing assigns a value
that depends only on the success probability, whereas feedback work values the
full side-information channel.  This distinction remains even before one
addresses the additional costs required to make either implementation fully
autonomous.

\section{Acknowledgments}

The author thanks Marcin Paw\l{}owski for useful discussion and comments and the inspiring idea to consider Szilard engine in my setup. ChatGPT 5.5 was helpful in polishing the manuscript and as an intellectual sparring partner during the development and refinement of ideas. It was also used to draw Figure 1 with the schematic representation of the two-particle Szilard engine.
This work is carried out under IRA Programme, project no. FENG.02.01-IP.05-0006/23, financed by the FENG program 2021-2027, Priority FENG.02, Measure FENG.02.01., with the support of the FNP.

\appendix
\section{Proof of Theorem~\ref{thm:extrema}}
\label{app:proof}

We give the full proof of the extremal result.  Let \(q=1-p\), with
\(q\in(0,1/2)\), and use the parametrization in Eq.~\eqref{eq:tparam}.  The
mutual information is Eq.~\eqref{eq:Ipt}.  The function is even in \(t\).

For \(x\in(0,1)\),
\begin{equation}
h_2''(x)
=
-\frac{1}{(\ln2)x(1-x)}.
\label{eq:h2second}
\end{equation}
Differentiating Eq.~\eqref{eq:Ipt} twice and simplifying gives
\begin{equation}
I_p''(t)
=
\frac{
(1-2q)^2\left[q(1-q)+3t^2\right]
}{
(\ln2)
(q^2-t^2)
(1-4t^2)
\left[(1-q)^2-t^2\right]
}.
\label{eq:Isecond}
\end{equation}
For \(q\in(0,1/2)\) and \(|t|<q\), every factor in the denominator is positive:
\[
q^2-t^2>0,
\qquad
1-4t^2>1-4q^2>0,
\]
and
\[
(1-q)^2-t^2
>
(1-q)^2-q^2
=
1-2q
>
0.
\]
The numerator of Eq.~\eqref{eq:Isecond} is also positive.  Hence
\[
I_p''(t)>0
\]
for every \(t\in(-q,q)\), so \(I_p\) is strictly convex.

Since \(I_p(t)\) is even and strictly convex, its unique minimum is at \(t=0\).
At this point,
\[
e_0=e_1=q=1-p,
\]
and
\begin{align}
I_p(0)
&=
h_2(1/2)-h_2(q)
\nonumber\\
&=
1-h_2(p).
\label{eq:Iminproof}
\end{align}

A strictly convex function on a compact interval attains its maximum at an
endpoint.  Because \(I_p\) is even, the two endpoint values are equal.  At
\(t=q\), one has \(e_0=2q\), \(e_1=0\), and
\begin{align}
I_p(q)
&=
h_2\!\left(\frac12+q\right)
-\frac12h_2(2q)
\nonumber\\
&=
h_2\!\left(p-\frac12\right)
-\frac12h_2(2p-1),
\label{eq:Imaxproof}
\end{align}
where we used \(h_2(x)=h_2(1-x)\).  The endpoint \(t=-q\) gives the second
\(Z\)-type channel.  Multiplication by \(\kB T\ln2\) proves
Eqs.~\eqref{eq:wmin} and \eqref{eq:wmax}.

At \(p=1/2\), the condition \(e_0+e_1=1\) makes the two rows of the channel
matrix identical, so \(I(X{:}G)=0\).  At \(p=1\), one has \(e_0=e_1=0\),
hence \(G=X\) and \(I(X{:}G)=1\).  This completes the proof.

\bibliographystyle{apsrev4-2}
\bibliography{refs}

\end{document}